# Phase matched nonlinear optics via patterning layered materials

**TAYLOR K. FRYETT,[1] ALAN ZHAN[2], AND ARKA MAJUMDAR[1,2,*]**

[1]*Electrical Engineering, University of Washington, Seattle, WA-98195*
[2]*Physics Department University of Washington, Seattle, WA-98195*
*\*Corresponding author: arka@uw.edu*

**The ease of integration coupled with large second-order nonlinear coefficient of atomically thin layered 2D materials presents a unique opportunity to realize second-order nonlinearity in silicon compatible integrated photonic system. However, the phase matching requirement for second-order nonlinear optical processes makes the nanophotonic design difficult. We show that by nano-patterning the 2D material, quasi-phase matching can be achieved. Such patterning based phase-matching could potentially compensate for inevitable fabrication errors and significantly simplify the design process of the nonlinear nano-photonic devices.**

Realizing low-power nonlinear optics in a scalable way is important for both fundamental scientific studies, such as quantum simulation using correlated photons [1], and for technological applications, such as efficient nonlinear frequency conversion of single photons for quantum communication [2] or optoelectronic information processing [3]. Second-order $\chi^{(2)}$ nonlinearity is particularly promising as this effect has a stronger dependence on cavity quality factors (Q) than $\chi^{(3)}$ nonlinearities, and can potentially reach few-photon nonlinear optics [4, 5]. However, efficient $\chi^{(2)}$ processes require the phase matching condition to be satisfied [6]. Several methods exist for achieving phase matching in macroscopic systems such as utilizing birefringence in KDP crystals [7], or quasi-phase matching in PPLN or GaAs [8-10]. In integrated nanophotonic cavities, the phase matching condition amounts to maximizing a spatial overlap integral between cavity modes at the fundamental and harmonic frequencies [11].

Most integrated nonlinear optical platforms utilize large resonators as they are easier to design and fabricate compared to compact nonlinear devices. For instance, in large integrated ring resonators, phase and frequency matching have been attained by making the effective mode-indices at the fundamental and the second-harmonic frequency identical [12]. This method is only effective for large devices, on the order of hundreds of wavelengths or larger and still has stringent fabrication tolerances. These problems become more pronounced in smaller wavelength-scale devices, which are attractive as nonlinear effects occur at lower powers due to reduced mode volume. The limitation on device size originates from the mixing of the electric field components (with respect to the nonlinear crystal basis) in wavelength-scale cavities. In the presence of such mixing it becomes increasingly difficult to simultaneously satisfy both the phase and frequency matching conditions. One approach to solve this problem is to design cavities using sophisticated optimization techniques [11], which is computationally intensive and often results in devices that are very difficult to fabricate.

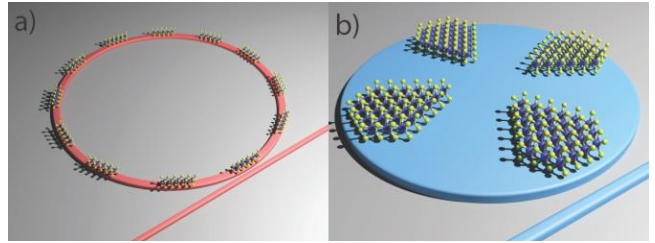

Figure 1: Schematic of (a) ring and (b) disk resonator with patterned layered materials on top for the phase-matching.

Atomically thin layered materials, in particular transition metal dichalcogenides (TMDCs) present a unique platform to create a hybrid nonlinear nano-photonic system [13]. Recently, researchers measured large $\chi^{(2)}$ coefficients in this material system [14, 15], and have integrated these materials with nanophotonic devices [16] to observe optically pumped lasing [17-19], strong exciton-photon coupling [20, 21], cavity enhanced electroluminescence, [22] and second harmonic generation [23-25]. For a perfectly phase-matched optical resonator, the effective nonlinearity in a layered 2D material integrated cavity is proportional to the product of the second order nonlinear coefficient and the thickness of the nonlinear material [13]. Due to the extreme thinness of 2D material the effective nonlinear interaction strength is reduced. However, this extreme thinness, and the evanescent nature of the interaction with the photonic device allow 2D materials to be integrated on a nano-cavity without significantly perturbing its electromagnetic modes. Via finite element electromagnetic simulations, we confirmed that the mismatch between the spatial distribution of modes with and without 2D materials on top is of the order of $10^{-4}$. For these simulations, we used graphene as the candidate 2D material for the availability of the model parameters, which have been proven to conform to experimental results[26]. We expect similar behavior for TMDC as well. Based on this intuition, we propose and theoretically analyze a method to obtain phase matching in ultra-compact nano-cavities by post-fabrication patterning of the 2D material. The efficiency of the second harmonic generation (SHG) in a patterned TMDC clad nano-cavity is evaluated by calculating the overlap integral. Our analysis suggests that, in a cavity without any phase matching, we can retain the modal overlap at ~65% of the value in a perfectly phase-matched cavity. Thus, for a nano-cavity with low modal overlap, by patterning the 2D materials we can satisfy the phase matching condition. Such ability is absent in the usual nonlinear cavities, where the cavity itself is usually made of the nonlinear material. Thus, patterning the nonlinear material completely changes the confined field. The robustness of the cavity modes against the perturbation of 2D material makes patterning based phase matching possible.

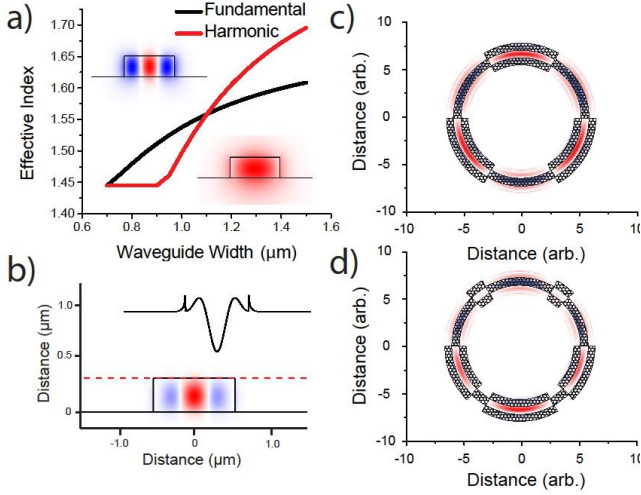

Figure 2: Layered material clad ring resonator: (a) effective mode indices in a large ring resonator as a function of the waveguide width. The upper inset corresponds to the mode profile of the harmonic mode whose effective index is plotted in red, while the lower inset is the fundamental mode, plotted in black. (b) The resulting nonlinear overlap from the fundamental and harmonic modes in (a). The red dotted line shows the position of the 2D material. The field experienced by the 2D material is shown in the top half of the figure. (c) The overlap surface profiles for a mode-matched ring, with the corresponding optimal 2D material shown by the hexagonal pattern. (d) The overlap and optimal pattern with a mode mismatch ($\Delta m = 2$).

The quantum limited conversion efficiency of SHG in a nano-cavity is determined by the spatial integral of the overlap between the fundamental and the second harmonic mode. A measure of the overlap integral is given by the $\beta$-factor [11], defined as:

$$|\beta|^2 = \left| \frac{1}{4} \frac{\iiint_{NL} dV \epsilon_0 \sum_{ijk} \chi^{(2)}_{ijk} \vec{E}^*_{fi} \vec{E}^*_{fj} \vec{E}_{hk}}{\iiint \epsilon |E_f|^2 dV \sqrt{\iiint \epsilon |E_h|^2 dV}} \right|^2 \quad (1)$$

Here $\chi^{(2)}_{ijk}$ are elements of the $\chi^{(2)}$ tensor in the Cartesian coordinates; $\vec{E}_f$ and $\vec{E}_h$ are the electric fields at the fundamental and second harmonic frequency, respectively; the integration in the numerator is performed only over the space where the nonlinear material is present, whereas the integral in the denominator is performed over the whole space. Unfortunately, this overlap integral often turns out to be very small due to the natural tendencies for the numerator to integrate to zero from mode symmetries. In nanophotonic cavities [11, 27], the spatial distribution of the nonlinear coefficient $\chi^{(2)}_{ijk}$ and the field distributions are interdependent complicating the optimization process. In a 2D material-clad nano-cavity we can engineer the spatial distribution of $\chi^{(2)}_{ijk}$ and cavity modes independently. Hence, the $\beta$-factor can be significantly changed by selectively patterning the 2D material, allowing the complex optimization problem to be split into two simpler optimization steps. To show the efficacy of patterning the 2D materials, here we calculate $\beta$ in 2D material ($\chi^{(2)}_{yyy} = 60 \, pm/V$) [15] clad nanophotonic cavities, such as ring and disk resonators (Fig. 1), and compare them to those calculated for previously reported cavity systems with aluminum nitride (AlN) with $\chi^{(2)}_{zzz} = 5$ pm/V [12] and gallium phosphide (GaP) with $\chi^{(2)}_{xyz} = 80 \, pm/V$ [28]. For these calculations, we assume fundamental mode at 1550nm and the second harmonic at 775nm. We note that, our methods can be employed for both travelling wave resonators and resonators with standing waves. While the mode-profiles in a travelling wave resonator change with time, the overlap integral is independent of time, and thus the 2D material patterning approach is valid.

For large rings, the optimization process for the bare cavity is less complex as there is little mixing between different polarizations. Hence, just ensuring same effective mode index $n_{eff}$ for the fundamental and second harmonic modes is sufficient. By satisfying the mode matching condition, we ensure that the harmonic and the square of the fundamental mode will retain the same relative phase as they propagate [6]. We calculate the $n_{eff}$ by using finite element methods, and determine the waveguide width where the effective index at 775 nm and 1550nm are same, keeping the waveguide thickness fixed at 350 nm. We find that the fifth order harmonic mode crosses the fundamental mode at a waveguide width of $\sim 1.1 \, \mu m$ (Fig. 2a). In a ring resonator, the waveguide bends cause a rotation of the electric field with respect to the monolayer crystal axes, forcing us to account for the tensor nature of the optical nonlinearity. TMDCs belong to the P6$_3$/mmc crystal symmetry group whose non-zero components of the $\chi^{(2)}$ tensor are $d_{yyy} = -d_{yxx} = -d_{xxy} = -d_{xyx}$ [14, 29]. Moreover, since the thickness of the monolayer is much smaller than the wavelengths we will take the fields to be roughly constant over its thickness and reduce the integral in the numerator to a surface integral multiplied by the 2D material interlayer thickness $d \approx 0.7$nm [15]. By converting to cylindrical coordinates and taking the radial position to be approximately constant over the radial integration, the expression for the overlap integral simplifies to:

$$|\beta|^2 = \left| \frac{\chi^{(2)}\epsilon_0 d}{4\sqrt{R(2\pi)^2}} \frac{\int_{NL} L_f^{*2} L_h dr \int_{NL} \sin(3\theta) d\theta}{\sqrt{\iint \epsilon |A_f|^2 da} \sqrt{\iint \epsilon |A_h|^2 da}} \right|^2 \quad (2)$$

Here $A_f$ and $A_h$ represent the modal profiles shown in Figs. 2a and 2b, while $L_f$ and $L_h$ correspond to the fields at the surface of the waveguide where the TMDC is placed; $R$ is the ring radius, $\epsilon$ is the permittivity, $\epsilon_0$ is the vacuum permittivity and $da$ is the area differential. Fig. 2b shows the nonlinear overlap of the cross-sectional mode profiles for the two previously discussed modes, with the inset plotting the overlap function for the cut designated by the dashed red line, where the TMDC will be placed. We find that the overlap integral could be increased by restricting the integration domain to only include portions of the overlap function with the same sign. Assuming that one of the 2D material axes is aligned along the polarization of the modes in the waveguide, a two- to three-fold increase in SHG power can be achieved in a straight waveguide, depending on which of the two patterning options (signs) is chosen. The integral over the azimuthal coordinate, $\theta$, (Eqn. 2) details how the overlap changes due to the tensor nature of second-order nonlinearity of the TMDCs. The odd nature of the integrand implies that without patterning, a perfectly circular ring with mode matching will nullify the second order nonlinearity of the TMDCs placed on top. However, these sign inversions can be nullified by inverting the radial patterning (Fig. 2c).

In cylindrically symmetric geometries the supported modes are discrete, allowing only momenta which create constructive interference over a round trip. Therefore, we can express the momentum mismatch as $\Delta m = m_h - 2m_f$ where $m_h$ and $m_f$ are the azimuthal mode numbers of the harmonic and fundamental modes, respectively. This leads to an additional azimuthal term, $\exp(i\Delta m\theta)$, inside the integral in the numerator of the overlap function, for a ring fabricated in nonlinear AlN and a ring with patterned 2D material on top. For the AlN ring, this implies a quenching of SHG light as the overlap function becomes identically zero. However, in the case of a 2D material integrated ring, this integral will essentially cause additional sign flips in the integrand that can be readily compensated for by adjusting the patterning scheme (Fig. 2d). We numerically calculate the $\beta$-factor for cavities with modes at the fundamental and the harmonic frequencies, but with a slight momentum mismatch. When the ring is phase-matched ($\Delta m = 0$), 2D material clad ring provides $\beta = 0.009 J^{-\frac{1}{2}}$, almost an order of magnitude smaller than that of AlN ring ($\beta = 0.089 J^{-1/2}$)[12]. However, for $\Delta m \neq 0$, the overlap intergral for the AlN ring identically falls to zero, whereas by appropriately designing the pattern, the overlap function remains nearly constant, at $\beta \approx 0.008 J^{-1/2}$, regardless of $\Delta m$ for the ring with patterned 2D materials. We note that, such nonzero momentum mismatch will inevitably arise from fabrication errors. By allowing the nonlinear overlap to be tailored after the fabrication of the resonator the tolerances can be significantly relaxed.

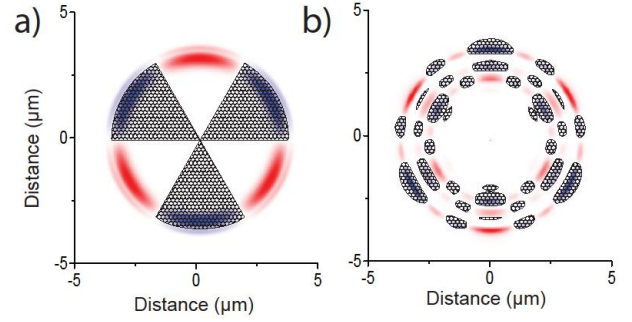

Figure 3: Nonlinear 2D material clad disk resonators: The nonlinear overlap integrand for a microdisk resonator integrated with 2D materials on top. The patterned hexagonal regions correspond the optimal patterns for the 2D material for (a) harmonic and fundamental modes with radial mode index $\rho = 1$, and $m_h = 2 \times m_f = 30$, and (b) $\rho_f = 1$, $\rho_h = 4$ with $m_f = 15$, and $m_h = 40$.

The performance enhancement due to the patterning is more prominent, when we consider the whispering gallery modes in a small resonator, where the polarizations are mixed with each other. In a small resonator, it is far more difficult to optimize the mode overlap function. This is underlined by the comparatively small number of experimental demonstration of an SHG device with modes at the fundamental and harmonic frequencies with good overlap [30]. One promising route to realize quasi phase matching (QPM) in micro-disk resonators made of III-V compounds is to exploit the 90-degree rotation about the (100) crystal axis, which is equivalent to a domain inversion. It has been demonstrated that QPM could be achieved by introducing a phase mismatch ($\Delta m = \pm 2$), that counteracted the domain inversions [31]. We compare the performance of our patterned 2D material approach with a GaP disk of radius $3\mu m$ and thickness of 200nm for the modes with azimuthal mode numbers 26 and 54 for the fundamental and second harmonic modes. With a value of $\chi^{(2)} = 80$ pm/V for GaP, we found $\beta = 4.734 J^{-1/2}$, compared to $\beta = 0.570 J^{-1/2}$ for patterned 2D materials on the same disk with the same modes. The bulk nonlinearity again demonstrates a larger overlap due to the increased thickness, but at the cost of much more stringent fabrication tolerances. To demonstrate the flexibility of patterned monolayer devices we design an arbitrary SiN disk resonator with a radius of 3 μm, and a height of 350 nm. By choosing $m_h = 40$ and $m_f = 15$ which correspond to modes at ~775 nm and ~1550 nm we find that $\beta$ can be has high as 0.1626 (Fig. 3a). The efficiency for such a mismatched GaP disk resonator drops by many orders of magnitude [32]. We note the $\beta$ values are larger for the small disk as expected from the small mode volume of the resonators. This overlap engineering can be extended to modes with more complicated spatial profiles as shown in Fig. 3b.

We report a new way to perform phase-matching using layered materials. The insensitivity of the cavity-confined spatial mode profiles to the 2D materials placed on top allows independent optimization of the nano-cavity and nonlinear medium. Such ability will significantly simplify the design process of nano-cavities, and can potentially circumvent the inevitable fabrication imperfections.

**Funding sources and acknowledgments.**

This work is supported by the National Science Foundation and the Air Force Office of Scientific Research-Young Investigator Program. The authors acknowledge useful discussion with Dr. Sonia Buckley.

**Funding.** National Science Foundation (NSF) (1433496, 1640986); Air Force Office of Scientific Research-Young Investigator Program (FA9550-15-1-0150).